\documentstyle[12pt]{article}
\input epsf.sty

\renewenvironment{thebibliography}[1]
 {\begin{list}{\arabic{enumi}.}
    {\usecounter{enumi} \setlength{\parsep}{0pt}
     \setlength{\itemsep}{3pt} \settowidth{\labelwidth}{#1.}
     \sloppy
    }}{\end{list}}

\begin{document}
\thispagestyle{empty}

\newlength{\captsize} \let\captsize=\small 

\font\fortssbx=cmssbx10 scaled \magstep2
\hbox to \hsize{
\raise.1in\hbox{\fortssbx Indiana University - Bloomington}
\hfill\vbox{\hbox{\bf IUHET-343}
            \hbox{September 1996}}
}

\vspace*{.5in}

\begin{center}
{\large\bf
Indirect Leptoquark Searches at Polarized Lepton 
Colliders
}\\[.1in]
\small
M.S.~Berger
\\[.1in]
\small\it
Physics Department, Indiana University, Bloomington, IN 47405,
USA\\
\end{center}

\vspace{.5in}

\begin{abstract}
We examine the utility of employing polarized lepton (electron and muon) beams
to perform indirect searches for scalar leptoquarks. 
We find that polarization can extend the reach in excluding leptoquark masses 
for both $e^+e^-$ and $\mu^+\mu^-$ machines. Polarization can also provide a 
diagnostic tool for determining leptoquark couplings.
\end{abstract}

\vspace{1.5in}

\noindent 
To appear in the {\it Proceedings of the 1996 DPF/DPB Summer Study on New 
Directions for High Energy Physics-Snowmass96}, Snowmass, CO, 25 June-12 July,
1996.

\newpage

\section{Introduction}
Theories attempting to unify the leptons and quarks in some common framework
often contain new states that couple to lepton-quark pairs, and hence are 
called leptoquarks\cite{guts}. Necessarily leptoquarks are color triplets,
carry both baryon number and lepton number, and can be either spin-0
(scalar) or spin-1 (vector) particles.
Perhaps the most well-known examples of leptoquarks appear as gauge
bosons of grand unified theories\cite{gg}. To prevent rapid proton decay they
must be very heavy and unobservable, or their couplings must be 
constrained by symmetries. Nonetheless, much work has been devoted 
to signals for the detection of leptoquarks at present and future 
colliders\cite{pp},\cite{ep},\cite{hr},\cite{ee},\cite{egam},\cite{gamgam}.
One potentially attractive source of light leptoquarks is in $E_6$ models 
where the scalar leptoquark can arise as the supersymmetric partner to the 
color-triplet quark that naturally resides in the fundamental representation
{\bf 27}. A recent review of the physics signals for leptoquarks can be found
in Ref.~\cite{review}.

At $e^+e^-$ and $\mu^+\mu^-$ colliders, pairs of leptoquarks can be produced 
directly via the $s$-channel $\gamma $ and $Z$ exchange. The reach for the 
leptoquark mass for this mode is essentially the kinematic limit, i.e.
$M_S< \sqrt{s}/2$. However even if a leptoquark is too massive to be produced 
directly, it can contribute\cite{hr},\cite{dreiner},\cite{choudhury} 
indirectly to the process 
$\ell^+\ell^-\to q\bar{q}$ by interfering with the Standard Model diagrams
as shown in Fig.~1. By examining the overall rate and the angular distribution,
indirect evidence for leptoquarks can be obtained. In this note, we examine
the bounds which can be placed on the leptoquark mass in this way, paying 
special attention to assessing the potential advantage that polarized electron 
or muon beams might provide.

\begin{center}
\epsfxsize=4.0in
\hspace*{0in}
\epsffile{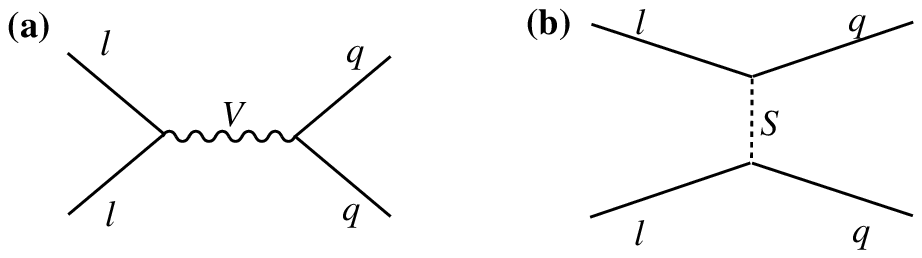}

\vspace*{0in}
\parbox{5.5in}{\small  Fig.1. The Feynman diagrams for the process 
$\ell^+\ell^-\to q\overline{q}$ include the (a) Standard Model diagrams
involving $s$-channel $V=\gamma,Z$ exchange,
and (b) the hypothetical $t$-channel leptoquark $S$ exchange.}
\end{center}

The polarization of the beams of a lepton collider can serve two purposes
in indirect leptoquark searches: 
(1) it can extend the reach of the indirect search by serving to enhance 
the fraction of initial leptons to which the 
leptoquark couples; (2) it can measure the left-handed and right-handed 
couplings of the leptoquark separately. 
Light leptoquarks (less than a few hundred GeV)
must also satisfy strong constraints from
flavor changing neutral current processes, so that leptoquarks must couple
to a single generation of quarks and leptons. 
For the leptoquarks that might be detected at the 
multi-TeV machines considered here, the constraints
from low energy processes do not
necessarily apply, since (as shown below) 
the reach in leptoquark mass can exceed even 
10~TeV, for which the FCNC effects should be very much suppressed.

\begin{center}
\epsfxsize=4.0in
\hspace*{0in}
\epsffile{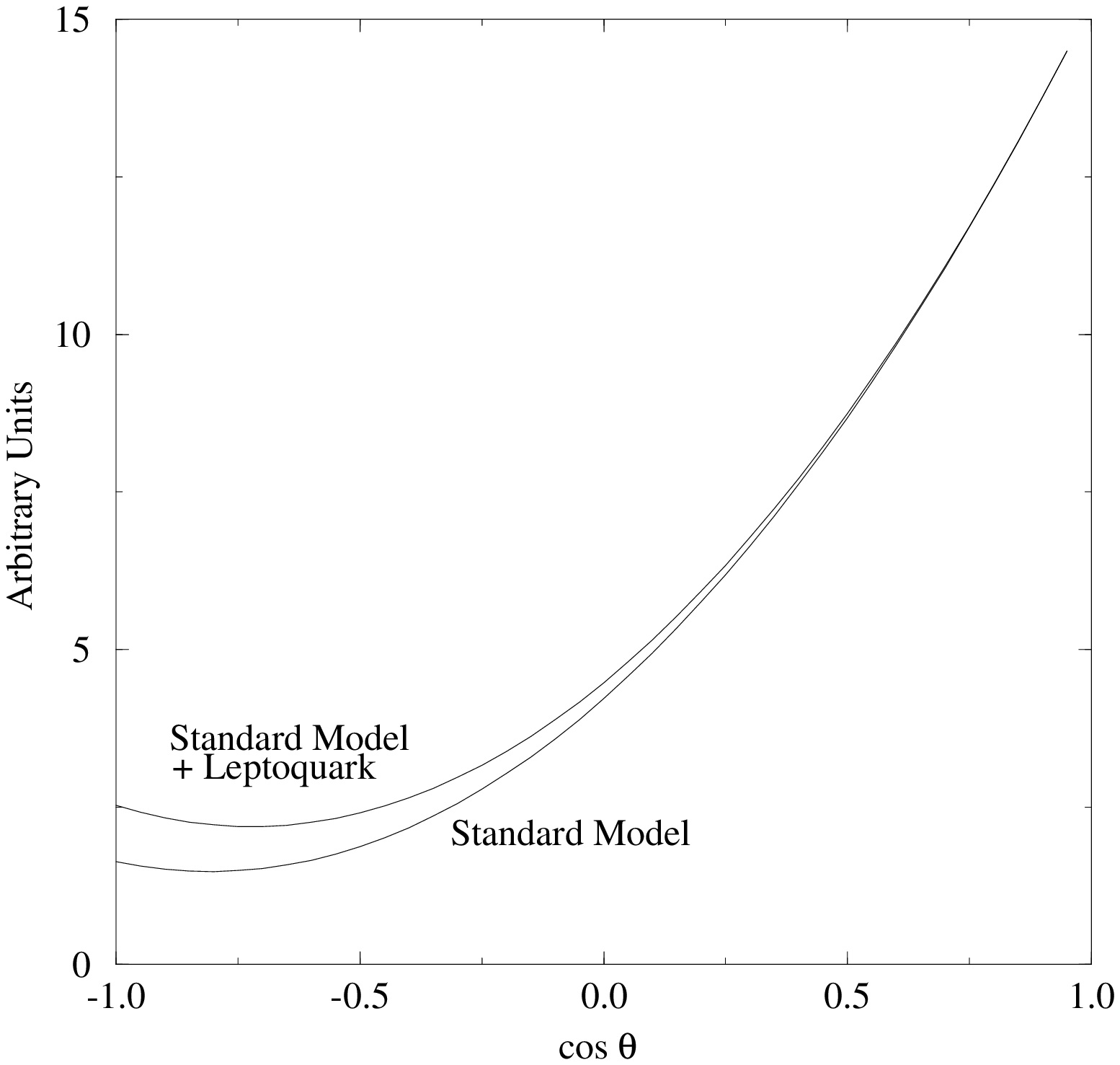}

\vspace*{-1.1in}
\parbox{5.5in}{\small  Fig.2. The angular distribution 
of $\ell^+\ell^-\to q\overline{q}$ in the Standard Model 
and including the effects of a scalar leptoquark for $M_S=8$~TeV and 
$\sqrt{s}=4$~TeV.}
\end{center}

The deviations from the Standard Model appear in the total cross section and 
the forward-backward asymmetry, $A_{FB}$\cite{hr}. 
In Fig.~2, the angular distribution
of the $q\overline{q}$ pair (the quark is taken to be $Q=2/3$)
is shown in the Standard Model and in the 
presence of a scalar leptoquark. The total cross section and $A_{FB}$ amount
to integrating this distribution in one or two bins respectively. In order to
maximize the sensitivity and following Choudhury\cite{choudhury}, we bin 
the cross section in 18 bins with $\Delta \cos \theta =0.1$ 
in the range $-0.9<\cos \theta < 0.9$ and perform
a $\chi ^2$-analysis to calculate the statistical significance of any 
deviations from the Standard Model. Therefore this procedure is simply a 
generalization of the measurement of the total cross section and $A_{FB}$.
The $\chi^2$ is determined in the usual way from the number of 
events expected in each bin in the Standard Model, $n_j^{\rm SM}$, and the 
number of events including the leptoquark, $n_j^{\rm LQ}$, expected in each 
bin, as
\begin{eqnarray}
\chi ^2&=&\sum _{j=1}^{18}{{(n_j^{\rm LQ}-n_j^{\rm SM})^2}
\over {n_j^{\rm SM}}}\;.
\end{eqnarray}

The additional piece in the Lagrangian that is of relevance to us can be
parametrized in the form
\begin{eqnarray}
{\cal L}&=&gS\bar{q}(\lambda_L P_L + \lambda_R P_R)\ell \;,
\end{eqnarray}
where $g$ is the weak coupling constant (to set the overall magnitude of the 
interaction) and $\lambda _{L,R}$ are dimensionless constants. $P_L$ and $P_R$ are the
left- and right-handed projectors. The size of the interference effect will be
determined by the three parameters $M_S$, $\lambda_L$ and $\lambda_R$.

Let us now concentrate on the interactions
$\ell^-(P^-) \ell^+(P^+) \rightarrow q\overline{q}$,
where the produced quark has $Q=2/3$. 
$P^-$ and $P^+$ are the polarization of the colliding leptons, and can be 
either left- or right-handed (we choose to define them such that they are 
always positive). 
The amplitudes for the diagrams presented in Fig.~1 have been presented for
the unpolarized case in Ref.~\cite{hr}, and is generalized to the case with
polarization in Ref.~\cite{choudhury}. So we do not repeat them here, and 
proceed directly to the results.

\section{Electron-Positron Collider}
The possibility of a multi-TeV $e^+e^-$ collider has begun to be taken 
seriously, and the 
physics potential of such a machine has started to be assessed.
It is expected that substantial polarization in the electron beam can 
be achieved, while the polarization of the positron beam might not be 
possible.
Figure~3 shows the 95\% c.l. bounds that could be achieved on a leptoquark with
right-handed couplings ($\lambda _L=0$) at a
$\sqrt{s}=4$~TeV $e^+e^-$ collider, with nonpolarized beams and with 
80\% and 100\% polarization of the electron beam. We have assumed 
integrated luminosity $L_0$  and efficiency $\epsilon$ for detecting the final 
state quarks so that $\epsilon L_0=70 {\rm fb}^{-1}$.
This reflects the luminosity benchmark of $L_0=100 {\rm fb}^{-1}$ and assumes
that the tagging efficiency for charm quarks might be as high as 70\% at
the machine.
Polarization from 80\% to 100\% roughly brackets
the range that might reasonably be achievable for the electron beam. The 
option of polarizing the electron beam is clearly very useful, as it can lead
to an increase in the bound by as much as a factor of two. Figure~4 shows 
the same bounds for the case where the leptoquark has left-handed couplings
($\lambda _R=0$). In this case the improvement is more modest but still 
nonnegligible.

In general a leptoquark would have both left- and right-handed couplings.
The bounds that can be achieved are
substantially larger than the collider energy, provided the leptoquark 
couplings are not too small compared to the weak coupling. 

\begin{center}
\epsfxsize=4.0in
\hspace*{0in}
\epsffile{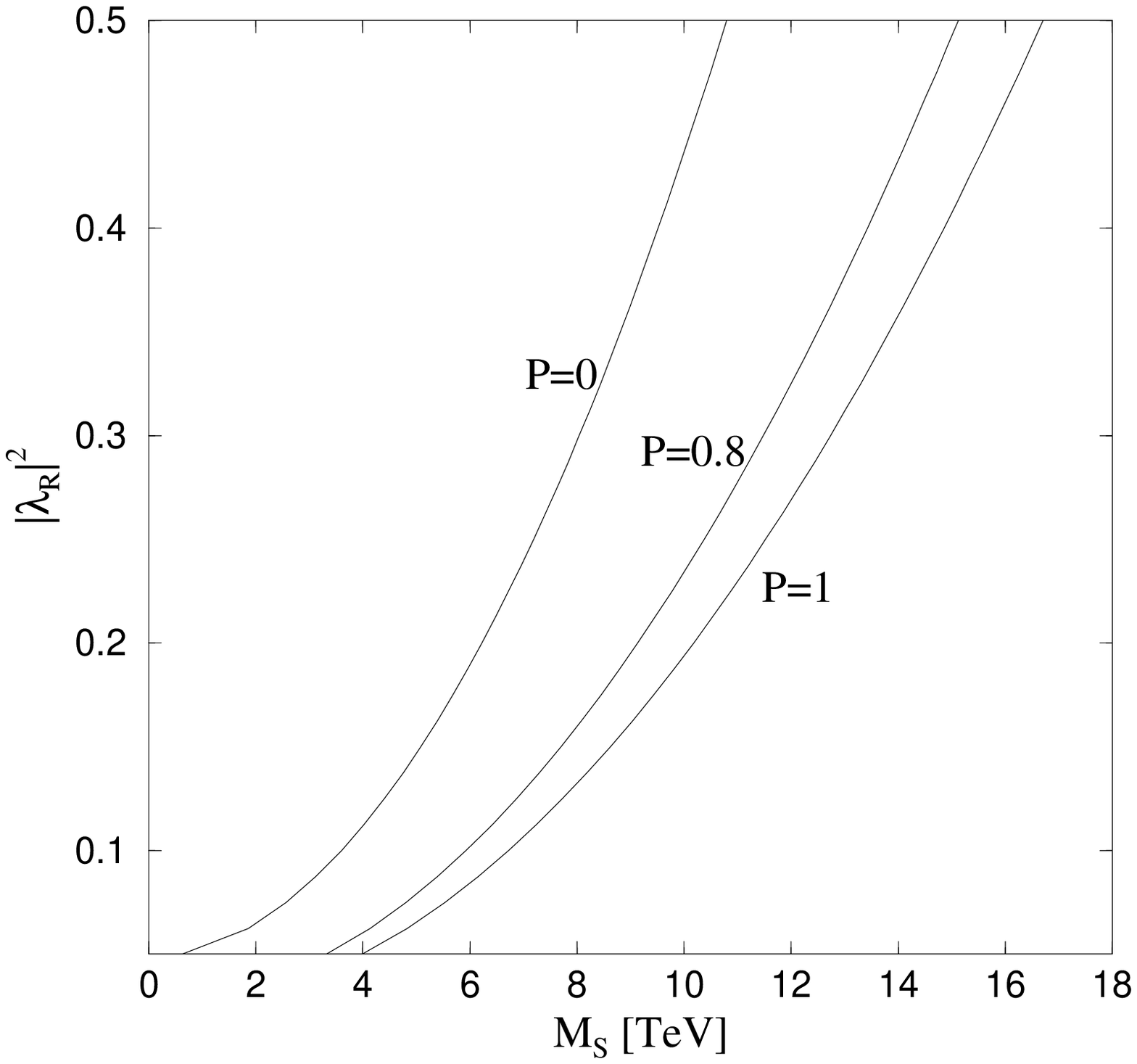}

\vspace*{-1.3in}
\parbox{5.5in}{\small  Fig.3. The 95\% c.l. bounds on leptoquark mass and 
couplings at 
a $\sqrt{s}=4$~TeV $e^+e^-$ collider for a leptoquark with right-handed 
couplings only ($\lambda _L=0$). The electron polarization $P$ is set to
0\%, 80\% and 100\%, and the positron is always unpolarized.
The area above each curve would be excluded.}
\end{center}

\begin{center}
\epsfxsize=4.0in
\hspace*{0in}
\epsffile{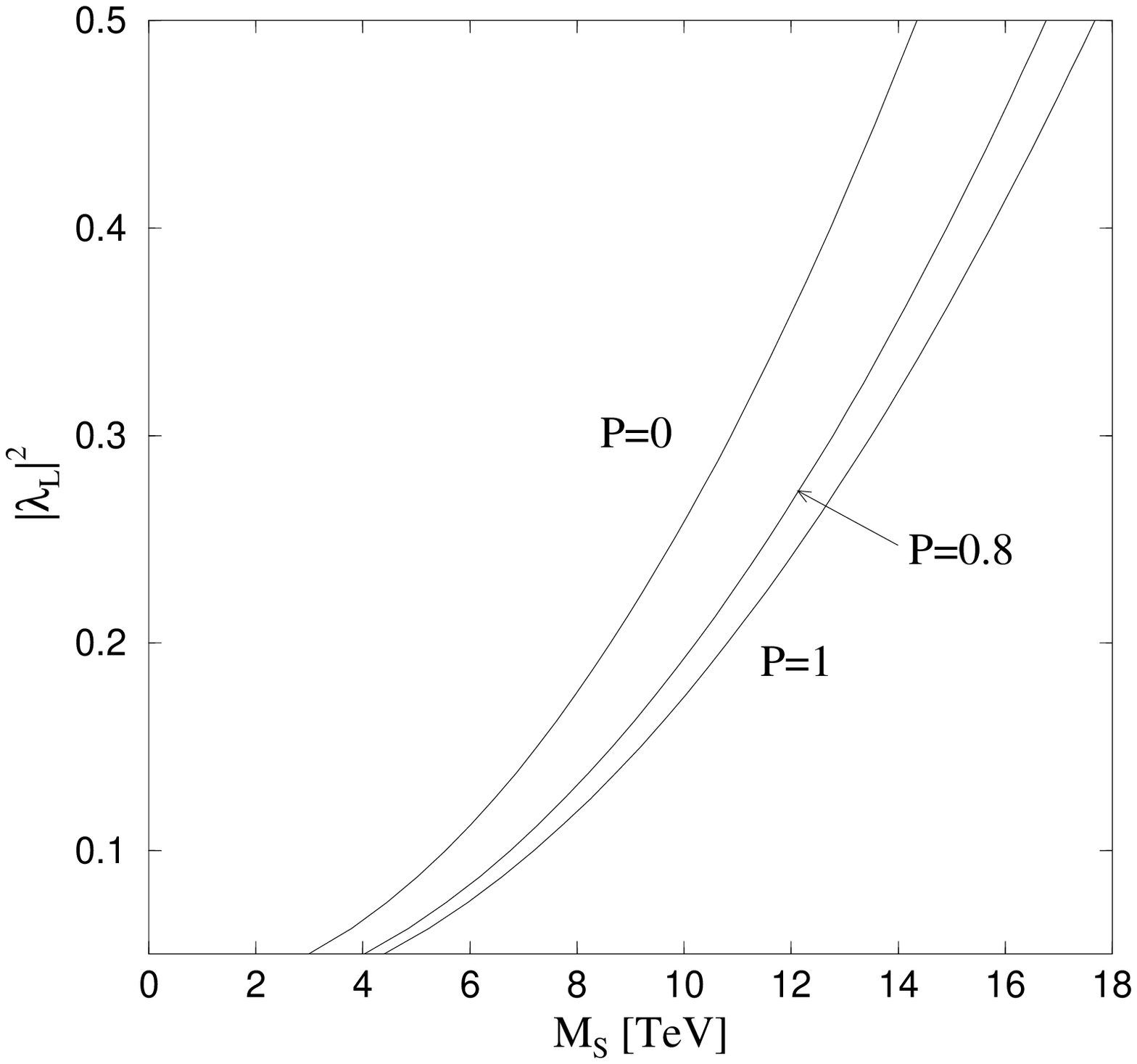}

\vspace*{-1.3in}
\parbox{5.5in}{\small  Fig.4. The 95\% c.l. bounds on leptoquark mass and 
couplings at 
a $\sqrt{s}=4$~TeV $e^+e^-$ collider for a leptoquark with left-handed 
couplings only ($\lambda _R=0$). The electron polarization $P$ is set to
0\%, 80\% and 100\%, and the positron is always unpolarized.
The area above each curve would be excluded.}
\end{center}

\section{Muon Collider}
There is increasing interest recently in the possible construction of a 
$\mu^+\mu^-$ collider\cite{mupmumi},\cite{saus},\cite{montauk},\cite{sfproc}.
The expectation is that a muon collider
with multi-TeV energy and the high luminosity
can be achieved\cite{neuffersaus,npsaus}.
Initial surveys of the physics potential of muon colliders have been carried
out\cite{workgr},\cite{sf}.
Both $\mu^+$ and $\mu^-$ beams can be at least partially polarized, but perhaps
with some loss of 
luminosity. At the Snowmass meeting 
a first study of the tradeoff between polarization and luminosity at a muon 
collider was
presented\cite{feas}. This analysis found that if one 
tolerates a drop in luminosity of a factor two, then one can achieve 
polarization of both beams at the level of $P^-=P^+=34\%$. 
(One could extend the 
polarization to 57\% with a reduction in the luminosity by a factor of eight.
This additional polarization does not prove useful if one must sacrifice so
much luminosity, at least for the leptoquark
searches studied here.)
It might be possible to maintain the luminosity at its full unpolarized value 
if the proton source intensity (a proton beam is used to create pions that
decay into muons for the collider) could be increased\cite{feas}.
We have chosen to present results for each of these three possible scenarios 
below.

\begin{center}
\epsfxsize=4.0in
\hspace*{0in}
\epsffile{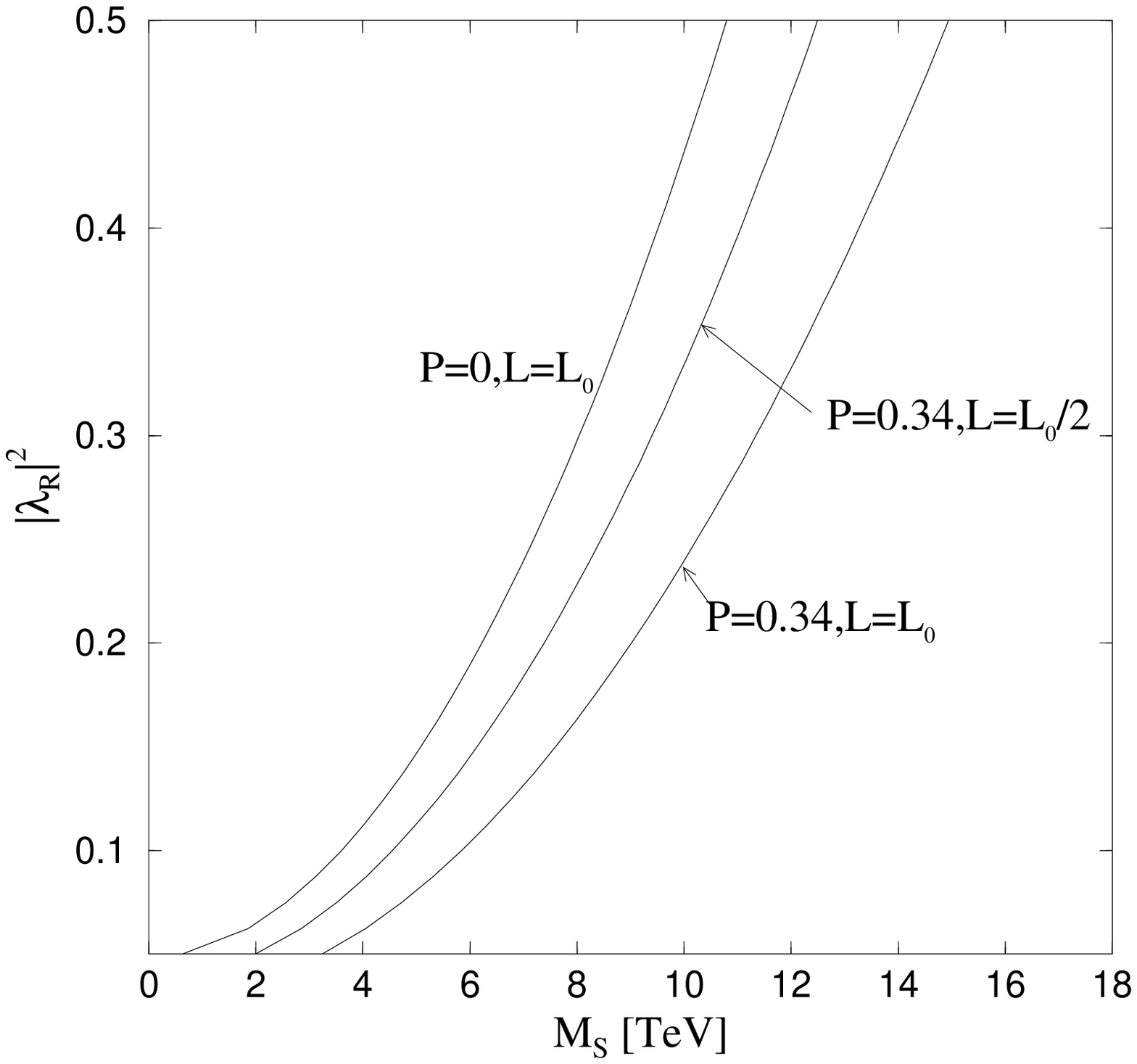}

\vspace*{-1.3in}
\parbox{5.5in}{\small  Fig.5. The 95\% c.l. bounds on leptoquark mass and 
couplings at 
a $\sqrt{s}=4$~TeV $\mu^+\mu^-$ collider for a leptoquark with right-handed 
couplings only ($\lambda _L=0$). The curves indicate the bounds for 
nonpolarized beams,  both $\mu^+$ and $\mu^-$ having 
polarization $P$ is set to 34\% and no reduction in luminosity, and 
both $\mu^+$ and $\mu^-$ having 
polarization $P$ is set to 34\% and a reduction in luminosity of 
a factor of two.
The area above each curve would be excluded.}
\end{center}

In Fig.~5 the 95\% c.l. bounds that can be obtained at a muon collider 
for a leptoquark
with right-handed couplings are shown for 
three cases: (1) unpolarized beams with integrated luminosity such that 
$\epsilon L_0=70{\rm fb}^{-1}$; (2) both
the $\mu^+$ and $\mu^-$ beams with 34\% polarization with the same luminosity
$L_0$; and (3) both the $\mu^+$ and $\mu^-$ beams with 34\% polarization but 
now including the expected reduction in luminosity
$L=L_0/2$. One sees that even with the reduction of luminosity one obtains 
improved bounds with polarized $\mu $ beams.  
In Fig.~6 the bounds that can be obtained at a muon collider for a leptoquark
with left-handed couplings are shown. In this case the expected luminosity 
reduction associated with polarizing the muon beams does not result in an 
improved bound. 

\begin{center}
\epsfxsize=4.0in
\hspace*{0in}
\epsffile{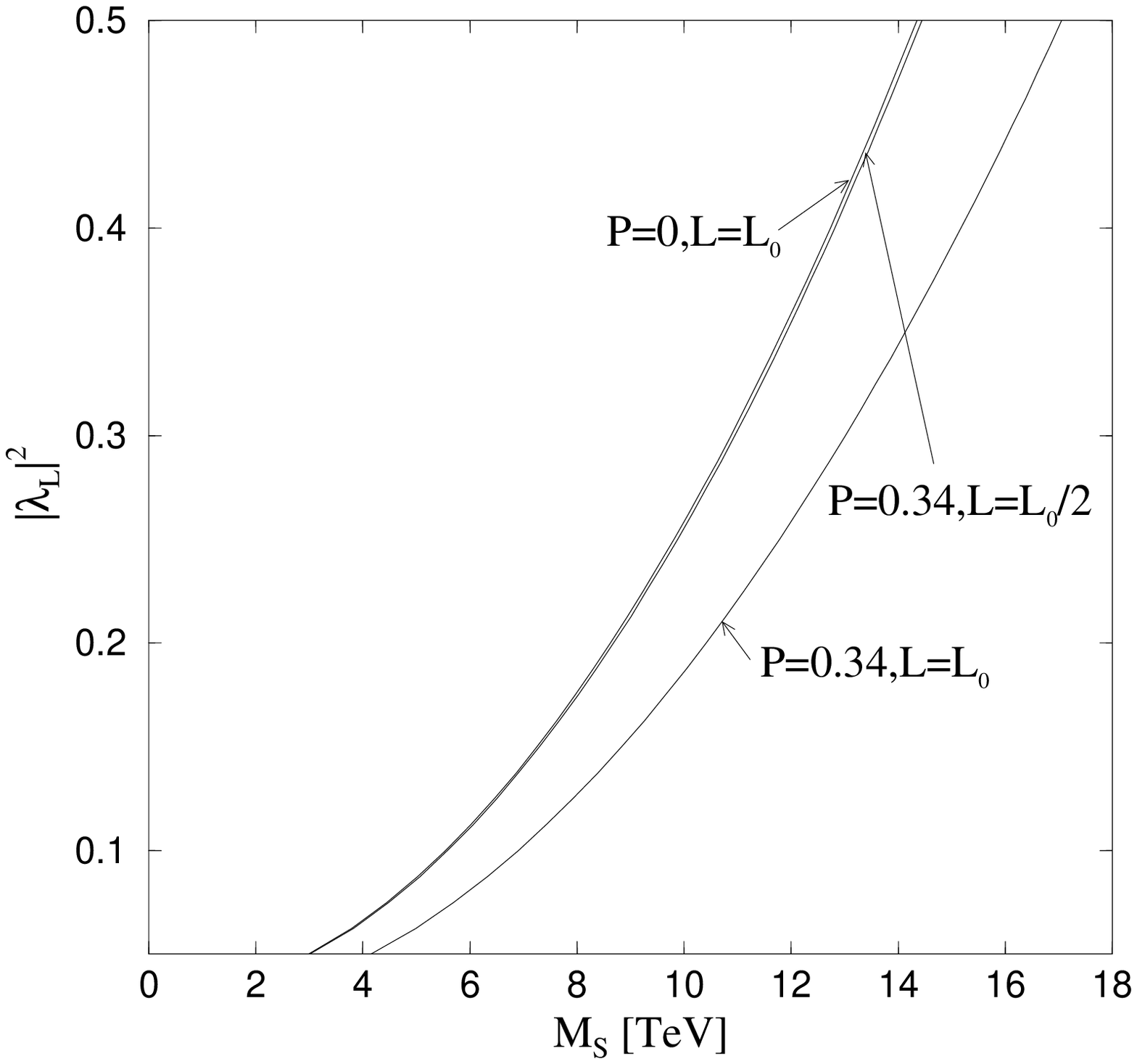}

\vspace*{-1.3in}
\parbox{5.5in}{\small  Fig.6. The 95\% c.l. bounds on leptoquark mass and 
couplings at 
a $\sqrt{s}=4$~TeV $\mu^+\mu^-$ collider for a leptoquark with left-handed 
couplings only ($\lambda _R=0$). The curves indicate the bounds for 
nonpolarized beams,  both $\mu^+$ and $\mu^-$ having 
polarization $P$ is set to 34\% and no reduction in luminosity, and 
both $\mu^+$ and $\mu^-$ having 
polarization $P$ is set to 34\% and a reduction in luminosity of 
a factor of two.
The area above each curve would be excluded.}
\end{center}

\section{Conclusions}

We have performed a first study of the indirect search for leptoquarks at 
multi-TeV lepton colliders. It is known already that polarization can be
advantageous at the NLC\cite{review},\cite{choudhury},
and we have shown by how much polarization 
is found to increase the lower bounds
on scalar leptoquark masses at both multi-TeV $e^+e^-$ machines and 
$\mu^+\mu^-$ machines. Of particular interest is the utility of 
polarization in the case of muon colliders, for which partial polarization of 
both beams is possible but comes at the cost of loss in luminosity. If one 
can achieve 34\% polarization in both muon beams, we find that this does 
improve the reach for leptoquarks if they couple to the right-handed muon, but
does not either improve or disimprove substantially the reach for leptoquarks
that couple to the left-handed muon. One should keep in mind that the 
expectations for the polarization and luminosity at a muon collider are very 
preliminary, and it might be possible to achieve polarization without 
significant reduction in the luminosity\cite{feas}. We find that 
polarizing the electron beam at an $e^+e^-$ collider improves the reach in 
scalar leptoquark mass, assuming no loss of luminosity.

Finally one can assess the utility of polarizing both beams as opposed to 
polarizing just one beam. This can be done by comparing Figs.~3 and 5 for the 
right-handed leptoquark case and Figs.~4 and 6 for the 
left-handed leptoquark case. We summarize the bound for leptoquarks with 
interactions of order the weak coupling strength in Table I, for both 
left-handed couplings ($|\lambda _L|^2=0.5,|\lambda _R|^2=0$) and right-handed 
couplings ($|\lambda _R|^2=0.5,|\lambda _L|^2=0$). The 95\% c.l. limits in the
two unknown parameter analysis, here translates into a 98.6\% c.l. when only 
the leptoquark mass is unknown.
For both cases one sees that the 34\% polarization
of both beams gives roughly the same bounds as a collider with one beam 
polarized at the 80-90\% level. 

\begin{table}[h]
\begin{center}
\caption{Bounds on leptoquark masses at 98.6\% 
confidence level, assuming either
left-handed couplings ($|\lambda _L|^2=0.5,|\lambda _R|^2=0$) or
right-handed couplings ($|\lambda _L|^2=0,|\lambda _R|^2=0.5$).}
\label{tab:sample}
\begin{tabular}{l|c|c}
\hline
\hline
Luminosity and & & \\
Polarization($\ell^-,\ell^+$)  & Coupling & $M_S$-Bound (TeV)\\
\hline
$L_0$ (0\%,0\%) & Left & 14.3  \\
            &  Right & 10.8  \\

\hline
$L_0$ (80\%,0\%) &  Left & 16.8  \\
            &  Right & 15.1  \\
\hline
$L_0$ (100\%,0\%) &  Left & 17.7  \\
            &  Right & 16.7  \\
\hline
$L_0$ (34\%,34\%)  & Left & 17.1  \\
                          & Right & 14.9  \\
\hline
$L_0/2$ (34\%,34\%)  & Left & 14.4  \\
                          & Right & 12.5  \\
\hline
\hline
\end{tabular}
\end{center}
\end{table}

It should be emphasized that there are many uses for polarization at these
machines, and the leptoquark search is just one entry on a long list of 
processes that should be studied to ascertain the full usefulness of including 
of polarization. Even without polarization we find the reach of a 4~TeV 
lepton collider is quite high: we find that leptoquarks
with couplings of roughly 
electroweak strength can be ruled out well above 10~TeV, and discovered even 
if they
have masses well above the collider energy.
Whether nature provides us with leptoquarks of about 10~TeV 
is, however, another matter indeed.

\section*{Acknowledgement}

I would like to thank J.L.~Hewett for suggesting this topic.
This work was supported in part by the U.S. Department of
Energy under Grant No. DE-FG02-95ER40661.

\end{document}